# Orthorhombic to tetragonal phase transition and superconductivity in the $Ba_2Cu_3O_4Cl_2$ compound

M. S. da Luz, C. A. M. dos Santos, B. Ferreira, and A. J. S. Machado

*Grupo de Supercondutividade, Departamento de Engenharia de Materiais,*

*FAENQUIL, 12.600-970, Lorena-SP, Brazil*

## ABSTRACT

In this work we have investigated the orthorhombic to tetragonal phase transition in the $Ba_2Cu_3O_4Cl_2$ compound. This transition was observed by X-ray powder diffractometry carried out in samples heat treated between 700 and 750$^O$C and also in samples with $Ba_2ZnCu_2O_4Cl_2$ composition. Results of X-ray diffractograms simulation confirm the phase transition. *dc*-Magnetization measurements performed in SQUID showed the existence of diamagnetism signal. The results suggest the existence of localized superconductivity and can explain the different magnetic properties reported in literature for the $Ba_2Cu_3O_4Cl_2$ compound.

**KEY-WORDS:** Oxychloride, structural transition, diamagnetism,

**Introduction**

The cooper oxyhalide compounds have received much attention in recent years because of their similar structures to the high critical temperature (high-$T_C$) superconducting oxides. The discovered of superconductivity in $Sr_2CuO_2F_{2+\delta}$ [1] stimulated further effort in the area and many new cooper oxyhalide systems have been prepared [2]. Among them, $Ba_2Cu_3O_4Cl_2$ (Ba2342) is one of the most interesting materials. Ba2342 seems to be an insulator with a tetragonal structure (space group: *I4/mmm*) and lattice constants $a = 5.517$ Å and $c = 13.808$ Å. This compound crystallizes in a layered structure composed of $Cu_3O_4$, Ba and Cl layers [3]. The $Cu_3O_4$ layers contain two types of Cu sites, $Cu_1$ and $Cu_{11}$ sites. The $Cu_1$ have an octahedral coordination with four square-planar oxygen ions in the basal plane, two chlorine ions at the apices, and the $Cu_{11}$ ions are surrounded only by four square planar oxygen ions. On the other hand, it is known that the $Cu_1$ spin is antiferromagnetically ordered and $Cu_{11}$ is paramagnetically ordered at room temperature [4,5]. The competition between these magnetic effects is considered to result in the multi-steps magnetic transitions in this compound. Furthermore, two magnetic transitions were observed in Ba2342; one at T~330K (denoted by $T_H$) and another at T~30K ($T_L$) [6, 7]. Between $T_H$ e $T_L$ the compound displays weak ferromagnetism. Noro *et al.* [8] studied the effect of the substitution of Cu for Ni and reported the disappearance of the weak ferromagnetic moment which decreased drastically by introducing a small amount of Ni as a dopant.

In contrast to the exposed above, diamagnetism and metal-like behavior have been observed for the Ba2342 compound at low temperatures, which suggest the existence of superconductivity in this compound as reported previously by some authors [9-12]. Another important aspect is that most of the high-Tc superconductors have crystalline structure containing Cu-O layers with CuO stoichiometry [13, 14]. In particular, the $YBa_2Cu_3O_{7-\delta}$ compound shows a phase transition from tetragonal to orthorhombic structure which occurs near 700°C in oxygen atmosphere [13,14]. The critical transition temperature depends on the oxygen partial pressure and occurs when the stoichiometry is near $YBa_2Cu_3O_{6.5}$. The highest superconducting transition temperature ($T_C$~90 K) is only observed for the orthorhombic structure [13, 14]. Also, high-pressure synthesis (at GPa) has been responsible to induce superconductivity in another oxychloride materials like $(Ca,K)_2CuO_2Cl_2$ ($T_C$ = 24 K) [15] and $(Ca, Na)_2CaCu_2O_4Cl_2$ ($T_C$ = 49 K) [16].

In this paper are reported results of X-ray powder diffractometry which suggest the existence of an orthorhombic to tetragonal transition in the

$Ba_2Cu_3O_4Cl_2$ compound. It is also shown results about the Zn doping on the properties of the $Ba_2Cu_3O_4Cl_2$. Magnetic measurements show diamagnetic signal which are in agreement with previous results suggesting the existence of localized superconductivity in this compound [9, 12, 17, 18].

1. **Experimental procedure**

Polycrystalline samples of $Ba_2Cu_3O_4Cl_2$ and $Ba_2ZnCu_2O_4Cl_2$ compositions were prepared by the solid state diffusion method using $BaCO_3$, $CuO$, $BaCl_2.2H_2O$ and $ZnO$ powders of high purity. The powders were mixed, ground, calcined at $700^OC$ for 24 hours, pressed into pellets, sintered at $750^OC$ for 48 h ($Ba_2Cu_3O_4Cl_2$) and 96 h ($Ba_2ZnCu_2O_4Cl_2$), and followed by a cooling to room temperature. Finally, the samples were heat treated at $700^OC$ ($Ba_2Cu_3O_4Cl_2$) and $730^OC$ ($Ba_2ZnCu_2O_4Cl_2$) for 48 h. The samples were characterized by x-ray powder diffractometry (Rich. Seifert - ISO Debyeflex 1001)) which were carefully indexed using the reference [19]. The x-ray diffraction intensity data of 2θ were colleted from 10 to 50°. A step-scan mode was adopted with a scanning step of 0.02° (in 2θ) and duration of 2s. The samples were also characterized by scanning electron microscopy (Jeol -

JXA840) with Energy Dispersive Spectrometry (EDS) in order to study the granular structures and grain compositions of the samples.

Using the Powder Cell software [20] to calculate lattice parameters and to simulate x-ray powder diffractograms (XRD) we were able to study the experimental x-ray diffractograms and the influence of the dopings and heat treatment on the crystalline structure of the $Ba_2Cu_3O_4Cl_2$ compound.

Magnetization measurements were performed using a Quantum Design Superconducting Quantum Interference device (SQUID) magnetometer.

## 2. Results and discussion

Figure 1 shows the XRD for the $Ba_2Cu_3O_4Cl_2$ sample heat treated at 750 (a) and 700$^O$C (b). Both diffractograms display single phases but a careful analysis of them show the existence of distinct peaks for the sample heat treated at 750$^O$C (see for example (103) reflections in the insets). This result is an evidence of a phase transition from orthorhombic to tetragonal in the $Ba_2Cu_3O_4Cl_2$ compound. In order to confirm this phase transition we performed simulation for the tetragonal and orthorhombic x-ray diffractograms. We adopted the crystallographic parameters shown in table 1 using the following lattice parameters: $a = b = 5.517$, $c = 13.808$ Å (space group: *I4/mmm*) and $a = 5.517$, $b = 5.550$ and $c = 13.808$ Å (space group: *Immm*) for the tetragonal and orthorhombic, respectively. The orthorhombic symmetry (*Immm*) was chosen because it resulted in the best agreement between experimental and simulated x-ray powder diffractograms. In the figure 2 one can observe that the result for the simulation of the tetragonal structure is in good agreement with the experimental x-ray powder diffractogram shown in the Fig. 1. For the orthorhombic structure we can see the appearance of double peaks (see for example the inset of this figure and also (200) reflections) which is similar to the sample heat treated at $700^{O}C$ (Fig. 1b). Thus, we can be used this method to study possible structural transition in the experimental x-ray powder diffractograms. A better comparison can be seen in the Fig. 3 which shows the (103) and (114) reflections for the tetragonal x-ray simulations related to the heat treatments at 750 and $700^{O}C$. Note the double peaks (see arrows) for the heat treatment at 700°C.

To study the influence of Zn additions on the $Ba_2Cu_3O_4Cl_2$ structure we have prepared samples with $Ba_2ZnCu_2O_4Cl_2$ composition. We are assuming that the Zn atoms can substitute Cu ions in the $Ba_2Cu_3O_4Cl_2$ structure as previously reported for other doping in layered cuprates [8, 21-23]. The x-ray powder diffraction of a sample, heat treated at 750°C, displays only peaks referred to $Ba_2Cu_3O_4Cl_2$ phase similarly to that shown in the figure 1(a). This result suggests that in this level of doping all Zn atoms are dissolved in the Ba2342 structure. In order to confirm if the Zn atoms were introduced in the Zn-doped sample, we carried out scanning electron microscopy. Figure 4(a) shows the scanning electron micrography of the $Ba_2ZnCu_2O_4Cl_2$ sample heat treated at $750^O$C. In the spot indicated by the arrow was performed an EDS analysis (displayed in figure 4b) which indicated that the Zn atoms were effectively incorporated inside the grains. This sample was also heat treated at $730^O$C. In the Figure 5 are shown the (103) and (114) reflections to these heat treatments. It is possible to observe the split of double peaks in the Zn-doped sample when treated at $730^O$C similar to non-doped sample heat treated at low temperature (see figure 1(b)).

In order to explore more details about the double peaks and the possible phase transition, in the figure 6 we show a comparison of the simulated (a) with the experimental peaks of the non-doped (b) and Zn-doped (c) samples. The comparison of the simulated and the experimental peaks for non-doped and Zn-doped samples demonstrate good agreement. The similarities between experimental and simulated diffractograms suggest that this oxychloride system has a phase transition from orthorhombic (*Immm*) to tetragonal (*I4/mmm*) structure. We have noted that the behaviors of x-ray powder diffractograms have some similarities with the $YBa_2Cu_3O_{7-\delta}$ ceramic superconductor which also shows a phase transition as a function of oxygen content [13, 14]. Thus, the phase transition observed in this work could explain the appearing of the diamagnetic signal and the different magnetic properties reported for the $Ba_2Cu_3O_4Cl_2$ compound [5-12].

Finally, Fig. 7 is displayed the temperature dependence of magnetic moment for a sample with $Ba_2Cu_3O_4Cl_2$ composition heat treated at 700°C (with orthorhombic symmetry). The presence of diamagnetism can be unambiguously observed in agreement with results reported previously [5,7-9]. Thus, our results are another indication of the existence of superconductivity in the $Ba_2Cu_3O_4Cl_2$ compound. We suggest that the orthorhombic phase can be responsible by diamagnetism behavior, but the superconductor volume in this system is small and probably it is related to the coexistence between orthorhombic-tetragonal phases. We speculate that the high pressure synthesis is able to stabilize the orthorhombic phase and promote the appearance of superconductivity in the $Ba_2Cu_3O_4Cl_2$ compound.

**Conclusion**

This work reports heat treatments and doping effects of Zn atoms on the $Ba_2Cu_3O_4Cl_2$ compound. Scanning electron microscopy coupled with EDS analysis demonstrated that the Zn atoms were incorporated inside grains of the Ba2324 compound. We have observed the formation of double peaks in the X-ray diffractograms of Zn-doped and non-doped samples heat treated at lower temperatures which were interpreted as a consequence of a structural transition in this oxychloride compound. Simulations of X-ray powder diffractograms confirm this idea and suggest that the phase transition is related to a changing from orthorhombic to tetragonal structure. *dc-magnetization* measurements performed in SQUID showed the existence of diamagnetism signal. The structural transition and the diamagnetism suggest the existence of superconductivity and can explain the different magnetic behaviors reported in literature in the $Ba_2Cu_3O_4Cl_2$ compound.

**Acknowledgements**

This work was partially supported by FAPESP (97/11020-8, 00/03610-4 and 00/08972-1). The authors would like to thank C. A. Rodrigues by EDS analysis.

**Table 1:** Crystallographic data for the $Ba_2Cu_3O_4Cl_2$ with tetragonal (I4/mmm) and orthorhombic (Immm) symmetries.

| | | | Atomic positions | | |
|---|---|---|---|---|---|
| Atom | Wyck (I4/mmm) | Wyck (Immm) | x | y | z |
| Ba | 4e | 4i | 0.000 | 0.000 | 0.365 |
| Cu1 | 4c | 2d | 0.000 | 0.500 | 0.000 |
| Cu2 | 2a | 2a | 0.000 | 0.000 | 0.000 |
| O | 4d | 4j | 0.000 | 0.500 | 0.250 |
| Cl | 8h | 4e | 0.250 | 0.000 | 0.000 |

**Table 2:** Lattice parameters for the $Ba_2Cu_3O_4Cl_2$ and $Ba_2ZnCu_2O_4Cl_2$ samples

| Sample | Heat treatment (°C) | Lattice Parameters (Å) | | | Structure |
|---|---|---|---|---|---|
| | | a | b | c | |
| $Ba_2Cu_3O_4Cl_2$ | 750 | 5.5129 | 5.5129 | 13.8216 | Tetragonal |
| | 700 | 5.4902 | 5.5249 | 13.7925 | Orthorhombic |
| $Ba_2ZnCu_2O_4Cl_2$ | 750 | 5.5135 | 5.5135 | 13.8200 | Tetragonal |
| | 730 | 5.4779 | 5.5331 | 13.8203 | Orthorhombic |

# Figure Captions

**Figure 1:** Experimental diffractograms for the sample with $Ba_2Cu_3O_4Cl_2$ composition heat treated at 750 (a) and 700°C (b). Insets show (103) reflections.

**Figure 2:** Simulated x-ray powder diffractograms for the $Ba_2Cu_3O_4Cl_2$ with tetragonal (a) and orthorhombic (b) symmetries. Insets highlight the (103) simulated reflections.

**Figure 3:** Comparison of the (103) and (114) simulated with tetragonal symmetry and experimental peaks for the sample with $Ba_2Cu_3O_4Cl_2$ composition heat treated at 700 and 750°C.

**Figure 4:** (a) Scanning electron micrograph of the $Ba_2ZnCu_2O_4Cl_2$ sample. In (b) is shown the EDS analysis in the grain indicated by the arrow.

**Figure 5:** (103) and (114) experimental reflections for the sample with Ba$_2$ZnCu$_2$O$_4$Cl$_2$ composition heat treated at 750 (above) and 730°C (below).

**Figure 6:** Comparison of the (103) experimental reflections for the Ba$_2$ZnCu$_2$O$_4$Cl$_2$ sample heat treated at 730 (a) and 700°C (b) with the (103) simulated peak for the orthorhombic symmetry (c).

**Figure 7:** Magnetic moment as a function of temperature for a Ba$_2$Cu$_3$O$_4$Cl$_2$ sample heat treated for 72h at 700°C.

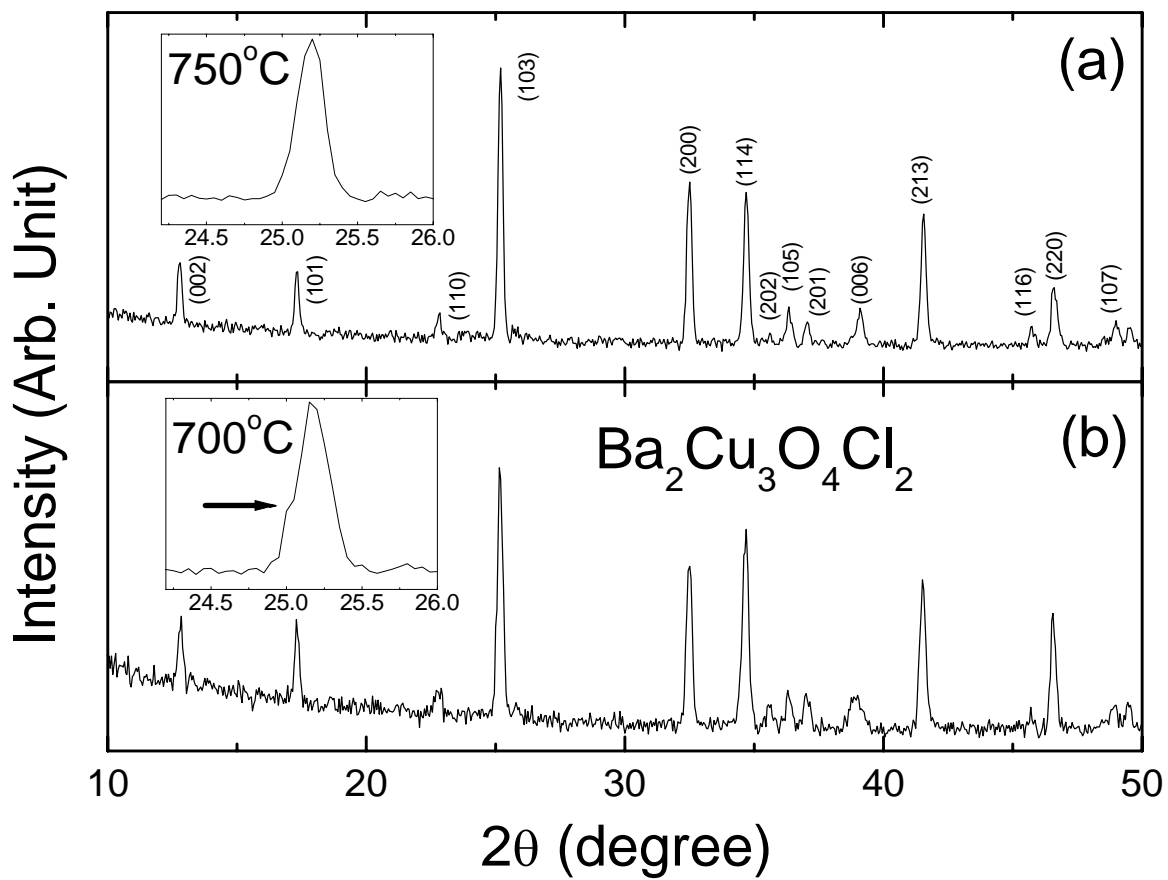

Figure 1 - M. S. da Luz et al.

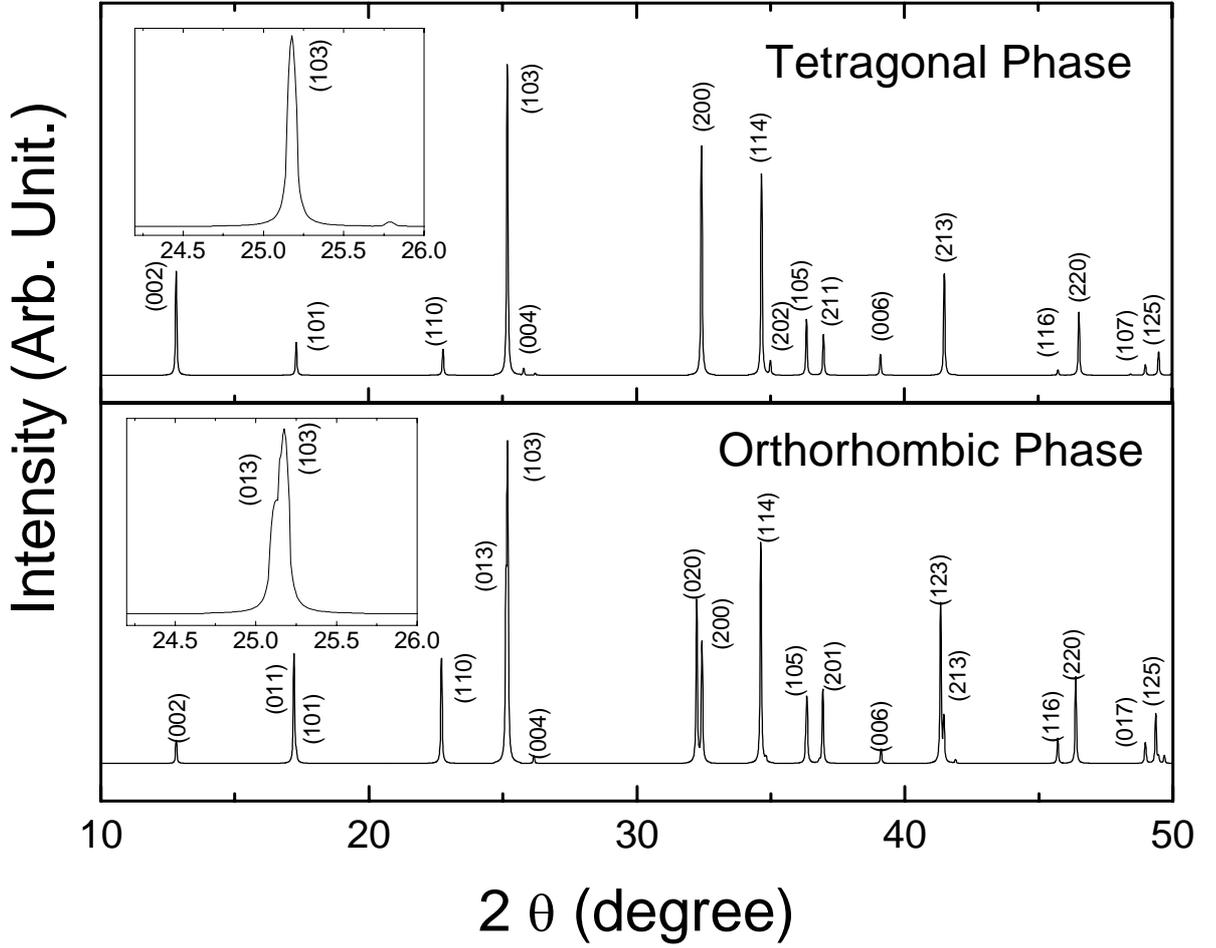

Figure 2 - M. S. da Luz *et al.*

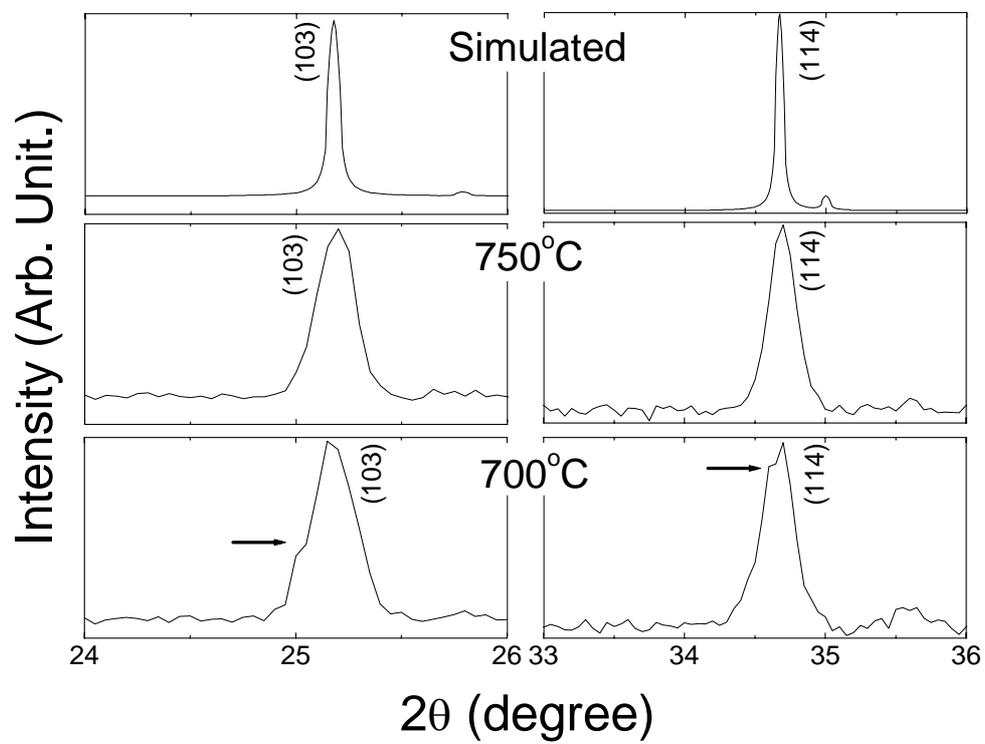

Figure 3 - M. S. da Luz *et al.*

Figure 4 – M. S. da Luz *et al.*

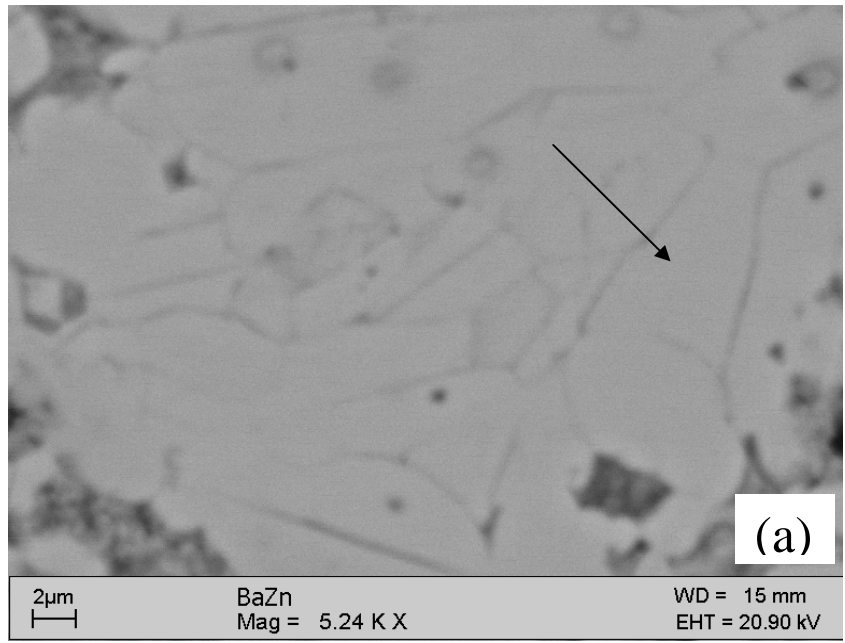

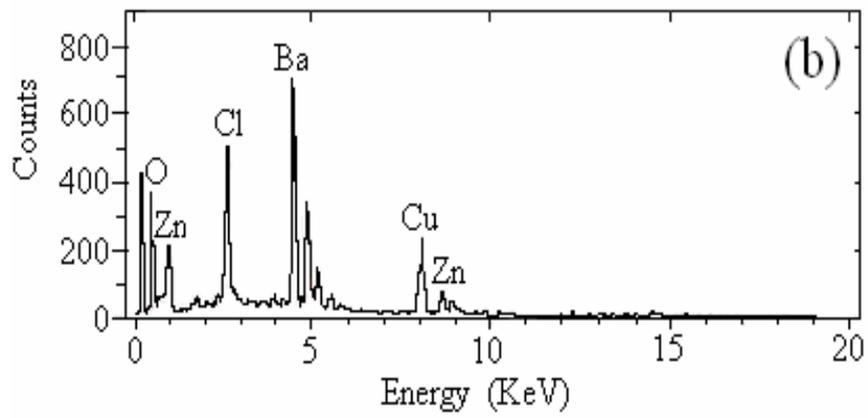

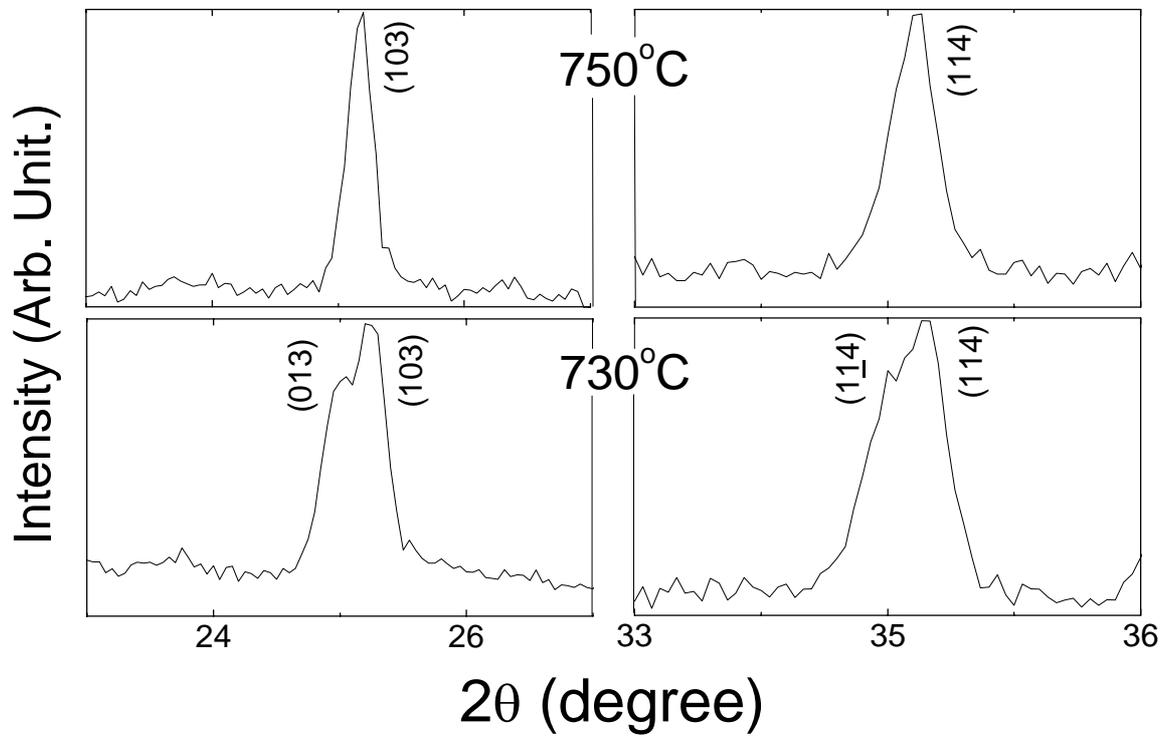

Figure 5 - M. S. da Luz *et al.*

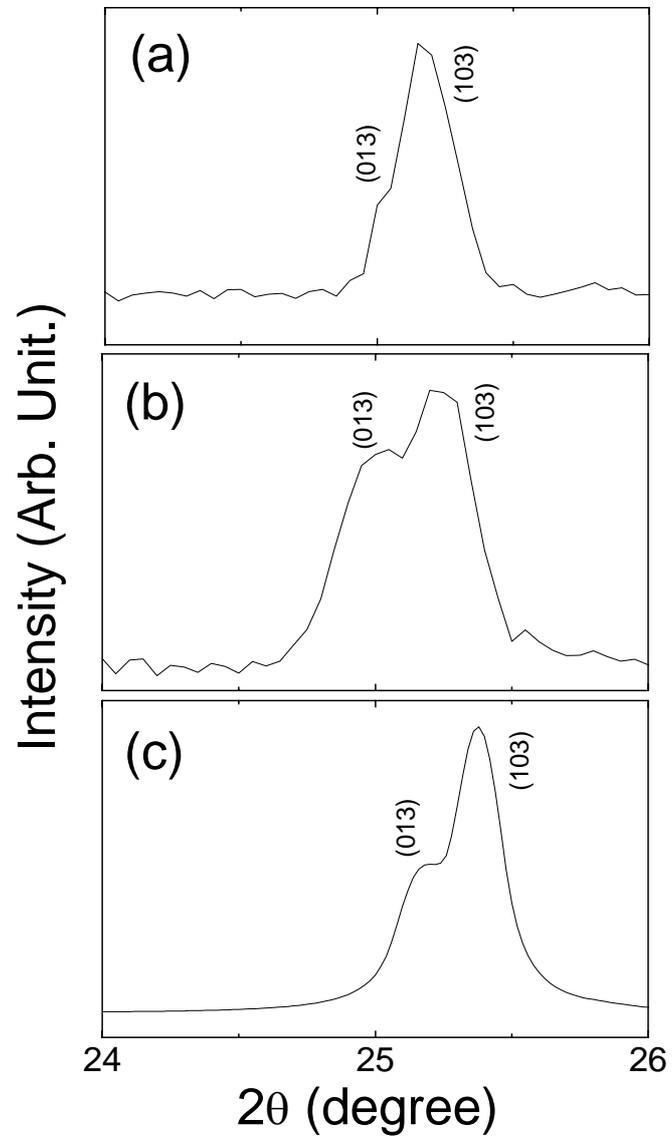

Figure 6 - M. S. da Luz *et al.*

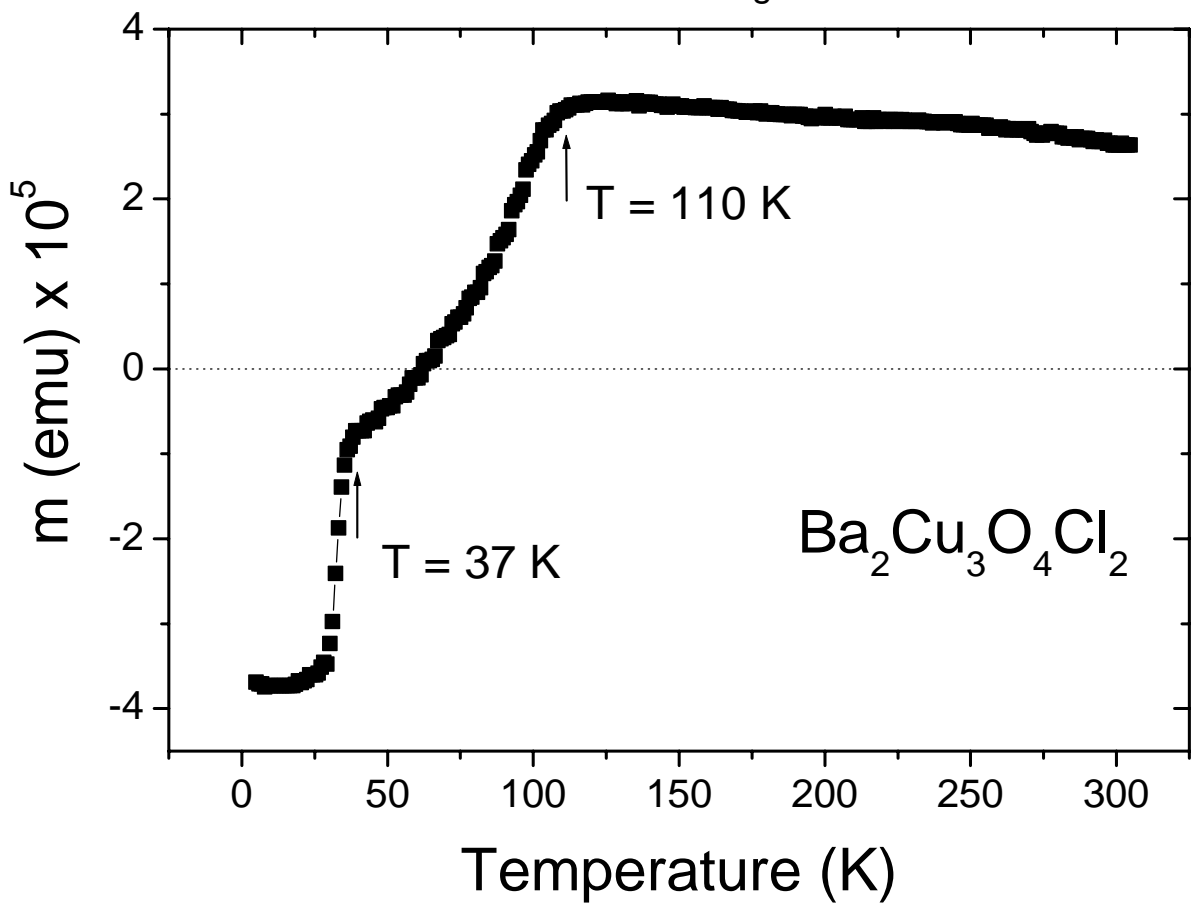

Figure 7 - M. S. da Luz *et al.*